\documentclass[conference]{IEEEtran}
\IEEEoverridecommandlockouts

\usepackage{cite}
\usepackage{amsmath,amssymb,amsfonts}
\usepackage{algorithmic}
\usepackage{graphicx}
\usepackage{textcomp}
\usepackage{xcolor}
\usepackage{physics} 
\usepackage{xcolor}
\usepackage{soul} 
\usepackage{amsmath} 

\def\BibTeX{{\rm B\kern-.05em{\sc i\kern-.025em b}\kern-.08em
    T\kern-.1667em\lower.7ex\hbox{E}\kern-.125emX}}
    
\begin{document}

\title{Near-field Beam Steering with Planar Antenna Array\\

\thanks{This research was funded by the Slovenian Research And Innovation Agency (ARIS) grant no. J2-4461, P2-0016, and PR-12348.}
}

\author{
    \IEEEauthorblockN{
        Aleš Simončič\IEEEauthorrefmark{1}\IEEEauthorrefmark{2},
        Andrej Hrovat\IEEEauthorrefmark{1}\IEEEauthorrefmark{2},
        Grega Morano\IEEEauthorrefmark{1}\IEEEauthorrefmark{2},
        Teodora Kocevska\IEEEauthorrefmark{1},
        Tomaž Javornik\IEEEauthorrefmark{1}\IEEEauthorrefmark{2}
    }
    \IEEEauthorblockA{
        \IEEEauthorrefmark{1}Department of Communication Systems, Jo\v{z}ef Stefan Institute, Ljubljana, Slovenia
    }
    \IEEEauthorblockA{
        \IEEEauthorrefmark{2}Jo\v{z}ef Stefan International Postgraduate School (IPS), Ljubljana, Slovenia.
    }
    \IEEEauthorblockA{
    Emails: \{ales.simoncic, andrej.hrovat, grega.morano, teodora.kocevska, tomaz.javornik\}@ijs.si
    }
}

\maketitle

\begin{abstract}

Beam steering enables manipulation of the electromagnetic radiation patterns in antenna array systems. 
A methodology for steering beams in the near field of a planar antenna array with known phase wavefront functions towards arbitrary azimuth and elevation angles is described in this paper. Rotation of the phase wavefront function is used while preserving the shape. The phase shifts for antenna element excitation currents are determined based on the distances from antenna elements to the nearest point on the rotated wavefront.
Beam steering utilizing a Gaussian and a Bessel beam is studied. Phase distribution examples for various steering directions are considered. For Bessel beam steering, a discussion on the resulting beam shapes and the steering impact on the polarization mismatch is provided. The results show a non-negligible magnitude of the cross-polarization with values depending on the steering direction.

\end{abstract}


\section{Introduction}

The deployment of extremely large antenna arrays in high-frequency bands to support the requirements of future wireless networks implies that wireless communications systems will operate in the near-field region different from the conventional systems typically operating in the far-field~\cite{liu24}.
The Fraunhofer distance distinguishes the near field from the far field regions for an arbitrary antenna, with the boundary between these regions being proportional to the square of the maximum dimensions of the antenna~\cite{balanis15b}.
In the context of an antenna array with a large number of antenna elements, typically, the largest dimension of each antenna element is significantly smaller than the largest dimension of the entire antenna array. Consequently, we infer that the far-field condition for each antenna element is much smaller than the far-field condition of the entire antenna array. This observation allows us to identify a region where antenna elements operate in the far field, thereby justifying the far-field approximations for modeling the emitted electromagnetic waves while the overall antenna array continues to operate in the near field. This presents an opportunity to exploit the antenna array's near field and employ wavefront engineering techniques to generate beams that would not be achievable in the far field~\cite{headland18}.

Through manipulation with the current excitation of the antenna elements in both magnitude and phase, the generation of various phase wavefronts representing the line that connects the points of a generated wave with same phase is possible, leading to the creation of diverse beams. In the literature, mostly the beam focusing, Bessel beams, Airy beams, and bottle beams, with the possibility of incorporating orthogonal angular momentum to the transmitted wave, have been studied~\cite{singh23,li18}. The beams are used in contemporary applications areas such as energy focusing, blockage mitigation, and multiple uncorrelated channels. 

In practical settings, such as in wireless communication systems or radar applications, with static antenna arrays and dynamic environments due to the movement of objects or changes in the propagation medium, the need for beam steering is emphasized. The use of fixed beams that can not adapt to environmental changes in real-time is inadequate. The ability to steer beams arbitrarily is essential for maximizing the performance and adaptability of antenna arrays.

In this paper, a methodology to steer the beam with known phase wavefront function in three-dimensional space is described. The polarization mismatch in the near field resulting from the beam steering process for a Bessel beam is addressed.

The structure of this paper is as follows. After the introduction, an overview of previous work is given in Section~\ref{Sec: related work}, and the system model is presented in Section~\ref{Sec: System model}. Section~\ref{Sec: Beam steering} presents the method for steering the beams with a known phase wavefront function.
Subsequently, in Section~\ref{Sec: Beam steering examples} Gaussian and Bessel beam steering is studied, along with phase distributions for various steering directions and presentation of the corresponding figures of steered beams.
The paper concludes with a summary and discussion.
For denoting the vectors boldface is used. The symbol $\mathbf{u}$ represents the unit vectors.

\section{Related work} \label{Sec: related work}

Significant advancements in wavefront engineering are reported in the literature, particularly in the domain of optics, where the insights and methodologies for tailored manipulation of electromagnetic waves enabled many applications ranging from imaging to laser beam shaping~\cite{yaras10,tsang16,arlt00}. The use of the available techniques in radio wave communication and sensing requires careful modification to consider the longer wavelength, hardware constraints, interaction with the propagation environment, and application requirements~\cite{jamshed20,pant23}.

In~\cite{singh22c}, Bessel beams were compared with Gaussian beams as representative of conventional far-field beamforming techniques and the numerical analysis was reported. The results highlight the advantages of Bessel beams, particularly in enhancing the resilience of communication links within (sub)THz frequency bands and near-field environments. The implementation of Gaussian beams and beam focusing techniques was studied, and the influence of the user position uncertainty on received signal strength was analysed in~\cite{acharya21}. 
The results show that beam focusing provides significant gain improvements and is sensitive to inaccuracies in the user position information attributed to the small focused area.

The use of wide bandwidth signals affects beam focusing due to the induced beam-splitting effect and causes alteration of the propagation range for Bessel beams as discussed in~\cite{singh23a}. 
To mitigate the adverse effects of the wide bandwidth on miss-focusing without resorting to the true time delays, a method based on the derivation of a frequency-modulated continuous wave (FMCW) as an optimal spatial phase modulation function was proposed in~\cite{myers22a}.

The polarization of the emitted signal from an antenna array has spatial variability in the near-field, contrary to the far-field scenario where a single polarization type is typically assumed. Consequently, the received signal strength fluctuates due to the movement of mobile users in front of the antenna array, although the polarization of the mobile user is constant.
The impact of polarization mismatch and the potential of its mitigation using phased arrays with dynamic polarization control was studied in~\cite{myers22}. Contrary to the spatially invariant optimal configuration observed in far-field scenarios, spatial variation across the array in the near field was shown. 

\section{System model} \label{Sec: System model}

\begin{figure}[!htb]
    \centering
    \includegraphics[width=1\columnwidth]{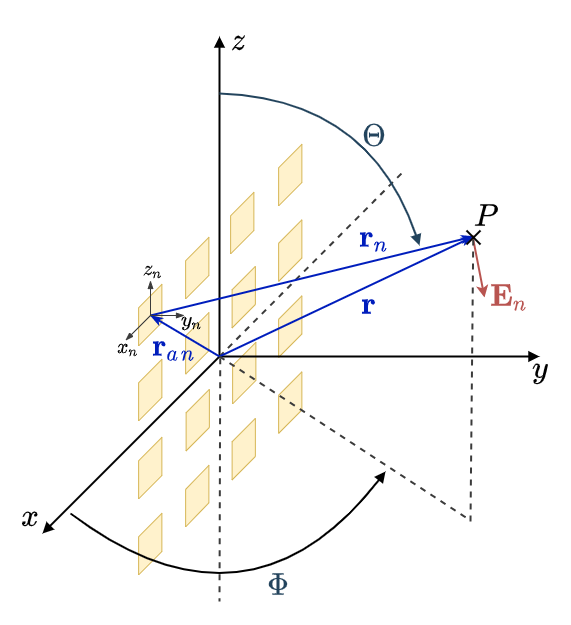}
    \caption{System model. The total electric field at point P contains the individual contributions of the antenna elements in the $xz$-plane.}
    \label{Fig:Antenna array in coordinate system}
\end{figure}

In this work, we consider a planar antenna array with $M$ antenna elements located in the $xz$-plane of a rectangular coordinate system, as illustrated in Figure~\ref{Fig:Antenna array in coordinate system} with yellow rectangles representing the array elements.
We denote $\mathbf{r_{an}}=(x_{an},y_{an},z_{an})$ as the vector from the origin of the coordinate system to $n$-th antenna element, where $n$ is the index of antenna element ranging from 1 to $M$. The vector $\mathbf{r}$ extends from the centre of the coordinate system to the point $P$ in which the electric field is calculated, while $\mathbf{r_{n}}$ represents the vector from the $n$-th antenna element to point $P$. The electric field in point $P$ from $n$-th antenna element is calculated as:
\begin{equation} 
     \mathbf{E_{n}} = \alpha \, I_{n} \frac{e^{-\jmath k \norm{\mathbf{r_{n}}} }}{\norm{\mathbf{r_{n}}}} \cdot \mathbf{u}_{\theta_{n}},
\end{equation}
where $\alpha$ is a proportionality constant set to unity given in $V/A$,
$I_{n}$ is the excitation current of the $n$-th antenna element in complex form, $k$ is the wavenumber, and $\mathbf{u}_{\theta_{n}}$ is the unit vector defining the polarization of the antenna element in the $\theta$ direction within the local coordinate system of the antenna element expressed as
 \begin{equation} 
     \mathbf{u}_{\theta_{n}} = ( \mathbf{u}_{x} \cos{\Phi_{n}} \cos{\theta_{n}} + \mathbf{u}_{y} \sin{\Phi_{n}} \cos{\theta_{n}} - \mathbf{u}_{z} \sin{\theta_{n}}),
\end{equation}
where $\Phi_{n}$ and $\theta_{n}$ are the azimuth and polar angles in local coordinates of $n$-th antenna element, respectively and
$\mathbf{u}_{x}$, $\mathbf{u}_{y}$, and $\mathbf{u}_{z}$ are the unit vectors in $x$, $y$, and $z$ direction, respectively. This model incorporates the polarization mismatch of emitted electromagnetic waves from antenna elements within the near field of the antenna array. We assume uniform excitation currents of equal magnitude and an isotropic radiation pattern for all antenna elements. Furthermore, we consider the transmitted signals to be monochromatic.

\section{Beam steering} \label{Sec: Beam steering}

\begin{figure}[!htb]
    \centering
    \includegraphics[width=1\columnwidth]{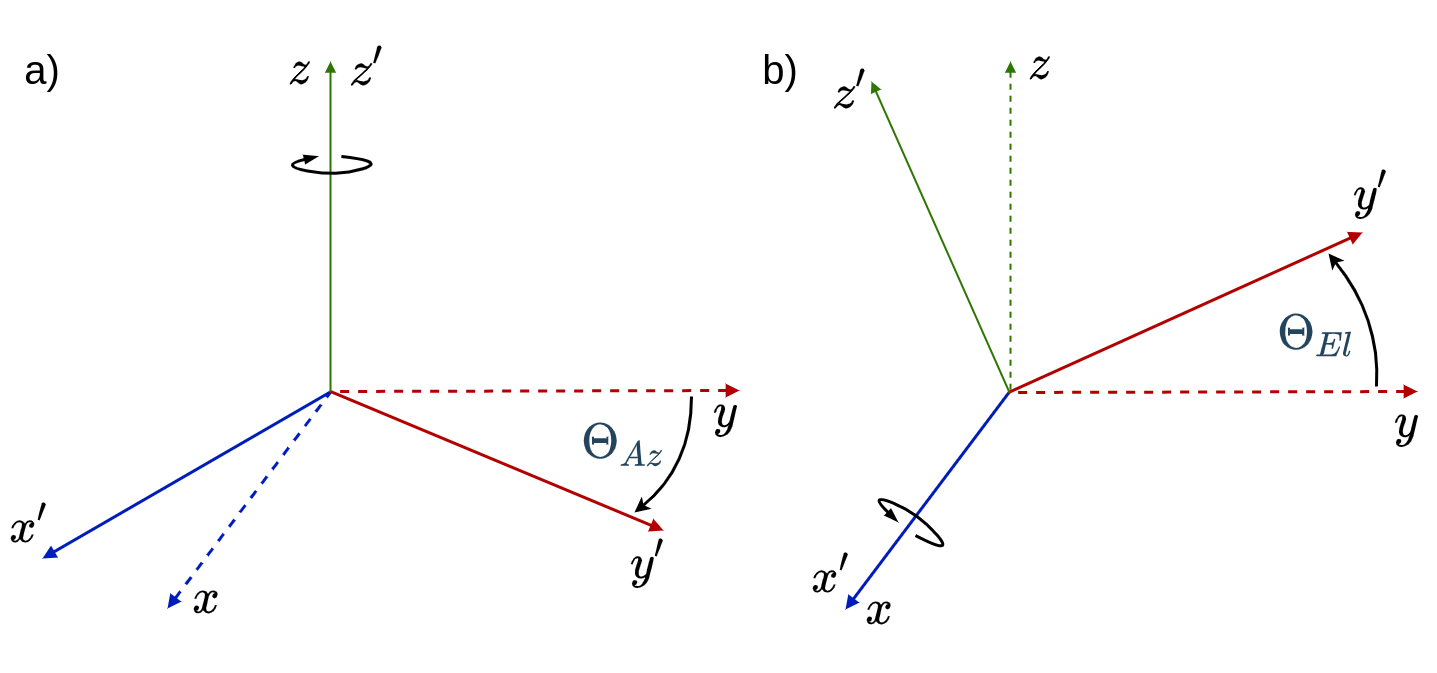}
    \caption{Rotation of coordinate system around $z$-axis (a) and $x$-axis (b).}
    \label{Fig:coordinate system transformation}
\end{figure}

In this section, we describe a method for steering the beams with a known function of the phase wavefront of the unsteered beam by calculating the phase shifts of the excitation current of the antenna elements such that the beam is steered in the arbitrary direction defined by azimuth ($\theta_{Az}$) and elevation ($\theta_{El}$) as shown in Figure~\ref{Fig:coordinate system transformation}(a) for the azimuth angle and Figure~\ref{Fig:coordinate system transformation}(b) for the elevation angle.

The first step is to determine the transformation matrix $R$ for the rotation of the original coordinate system by $\theta_{Az}$ around the $z$-axis and by $\theta_{El}$ around the $x$-axis, as defined by the rotation arrows in Figure~\ref{Fig:coordinate system transformation}, so that the $\mathbf{u}_{y'}$ direction corresponds to the desired steering direction. The transformation matrix $R$ is equal to matrix multiplication of transformation matrix $R_{x} (\theta_{El})$ which rotates the coordinate system around $x$-axis by $\theta_{El}$ and $R_{z} (\theta_{Az})$ which rotates the coordinate system around $z$-axis by $\theta_{Az}$:

\begin{equation} 
     R = R_{x} (\theta_{El}) \cdot R_{z} (\theta_{Az}).
\end{equation}
$R_{x} (\theta_{El})$ is defined as
\begin{equation}  
    R_{x} (\theta_{El}) = 
     \begin{bmatrix}
     1 & 0 & 0 \\
     0 & \cos{\theta_{El}} & -\sin{\theta_{El}} \\
     0 & \sin{\theta_{El}} & \cos{\theta_{El}}
     \end{bmatrix},
\end{equation}
and $R_{z} (\theta_{Az})$ is defined as
\begin{equation}  
    R_{z} (\theta_{Az}) = 
     \begin{bmatrix}
     \cos{\theta_{Az}} & \sin{\theta_{Az}} & 0 \\
     -\sin{\theta_{Az}} & \cos{\theta_{Az}} & 0 \\
     0 & 0 & 1
     \end{bmatrix}.
\end{equation}
The new coordinates are
\begin{equation}   \label{Eq:coordinates_transformation}
     \begin{bmatrix}
     x'\\
     y'\\
     z'
     \end{bmatrix} = R \cdot
     \begin{bmatrix}
     x\\
     y\\
     z
     \end{bmatrix}.
\end{equation}

The primary concept underlying the steering mechanism involves utilizing the wavefront function of an unsteered beam in a rotated or transformed coordinate system, where the direction of $\mathbf{u}_{y'}$ aligns with the desired steering direction. As the wavefront function retains its shape and solely undergoes spatial rotation, the resultant beam achieves effective steering while maintaining its inherent characteristics. Initially, we employ the unsteered wavefront function in the transformed coordinate system $y'=f(x', z')$. To express the same wavefront within the original coordinate system, we utilize the coordinate transformation equation delineated by (\ref{Eq:coordinates_transformation}). Consequently, the obtained phase wavefront function in the original coordinate system is denoted as $y_0 = f(x,z,\theta_{Az},\theta_{El})$.

\begin{figure}[!htb]
    \centering
    \includegraphics[width=1\columnwidth]{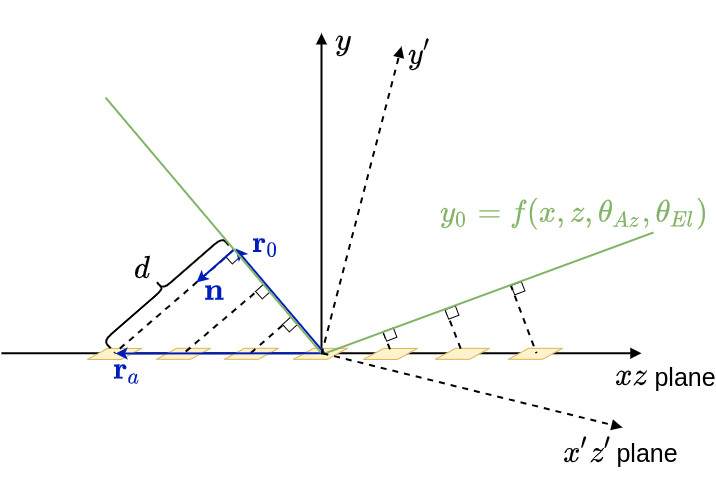}
    \caption{Minimal 
distance between an antenna element and the phase wavefront.}
    \label{Fig:Orthogonal lines: phase wavefronts - antennas}
\end{figure}

The subsequent stage involves calculating the minimum distances between each antenna element positioned in the $xz$-plane and the phase wavefront function defined in the original coordinate system. This task requires determining the length of a line originating from the location of the antenna element and ending at the intersection with the phase wavefront, under the condition that the line is perpendicular to the phase wavefront, as shown in Figure~\ref{Fig:Orthogonal lines: phase wavefronts - antennas}, where the minimum distance is denoted as $d$ for an antenna element taken as an example. The equation governing such a line's characteristics can be expressed as follows:
\begin{equation} \label{Eq:orthogonal line}
\mathbf{r_{0}} = \mathbf{r_{an}} - t \cdot \mathbf{n},
\end{equation}
where $\mathbf{r_{0}}$ denotes the coordinates on the surface represented by the phase wavefronts in the original coordinate system, defined as $(x,y_0(x,z,\theta_{Az},\theta_{El}),z)$, $\mathbf{n}$ represents the vector perpendicular to the phase wavefront, and $t$ is a constant value proportional to the length of the line.

The perpendicular vectors $\mathbf{n}$ to the wavefront surface $y_0(x,z,\theta_{Az},\theta_{El})$ are computed as:
\begin{equation} 
\mathbf{n} = \nabla (y_{0}-y) =  ( \frac{\partial y_{0}}{\partial x}, -1, \frac{\partial y_{0}}{\partial z} ).
\end{equation}

Expanding the expression of (\ref{Eq:orthogonal line}) while taking into account the aforementioned equalities yields:
\begin{equation} \label{Eq:orthogonal line system of equations} 
(x,y_0(x,z,\theta_{Az},\theta_{El}),z) = (x_{an}, y_{an}, z_{an}) - t \cdot ( \frac{\partial y_{0}}{\partial x}, -1, \frac{\partial y_{0}}{\partial z} ).
\end{equation}
The foregoing expression constitutes a system of three equations with three unknowns ($x$, $z$, $t$). By solving this system, we can ascertain $x_i$ and $z_i$, which correspond to the coordinates where the line originating from the antenna element's position intersects the phase wavefront. The point of intersection is $(x_i,y_0(x_i,z_i,\theta_{Az},\theta_{El}),z_i)$.
The distance $d$ is equal to the Euclidean distance between antenna element coordinates and the point of intersection:
\begin{equation} \label{Eq:d}
d_{an} = \sqrt{(x_{i}-x_{an})^{2} + y_0(x_i,z_i,\theta_{Az},\theta_{El})^{2} +(z_{i}-z_{an})^{2}}.
\end{equation}
The phase progression induced in the transmitted wave due to the distance $d$ traveled from antenna element $n$ is determined as follows:
\begin{equation} \label{Eq:antennas_phase_shifts}
\Delta \Phi_{an} (x_{an}, z_{an}) = \frac{2 \pi d_{an}}{\lambda},
\end{equation}
where $\lambda$ is the wavelength of the transmitted signal.
The excitation required for each antenna element to counteract the influence of phase progression as the signal travels towards the phase wavefront, thereby producing the steered beam, is expressed as $I_{n} = e^{\jmath \Delta \Phi_{an}}$. A depiction of the phase shifts in excitation corresponding to the $x$ and $z$ coordinates is commonly referred to as the phase distribution.

It is worth noting that the distance $d$ corresponds to the shortest distance between an antenna element and the phase wavefront. Therefore, an alternative approach to calculating the distance $d$ involves minimizing the squared distance function between two points (the antenna element coordinates and the closest point on the phase wavefront) denoted as 
\begin{equation}
f_{d} = (x-x_{an})^{2} + y_0(x,z,\theta_{Az},\theta_{El})^{2} +(z-z_{an})^{2},
\end{equation}
by solving for the unknown $x$ and $z$ coordinates through the system of two equations
\begin{equation}
\frac{\partial f_{d}}{\partial x} = 0, \,\,\, \frac{\partial f_{d}}{\partial z} = 0.
\end{equation}
A similar optimization methodology involves the utilization of Lagrange multipliers.
Despite yielding identical results, we observed that the method involving the computation of normal vectors on the phase wavefront surface is more efficient. Therefore, we focus our subsequent analysis on this method.

\section{Beam steering examples} \label{Sec: Beam steering examples}

This section describes the examples of steering Gaussian and Bessel beams~\cite{durnin87} using the method described above. We begin by describing the steering of Gaussian beams since the result corresponds to beam steering in the far field, which is well-studied and serves as a reference for the described steering method. Next, the steering of the Bessel beam is described as a representative beam in the near field, also addressing the lack of studies on how the steering affects the beam shape. The simulations for both beams were performed with a grid of 100x100 antenna elements operating at a frequency of 100 GHz with a spacing between the antenna elements of $\lambda /2$.

\subsection{Gaussian beam steering}

\begin{figure}[!htb]
    \centering
    \includegraphics[width=1\columnwidth]{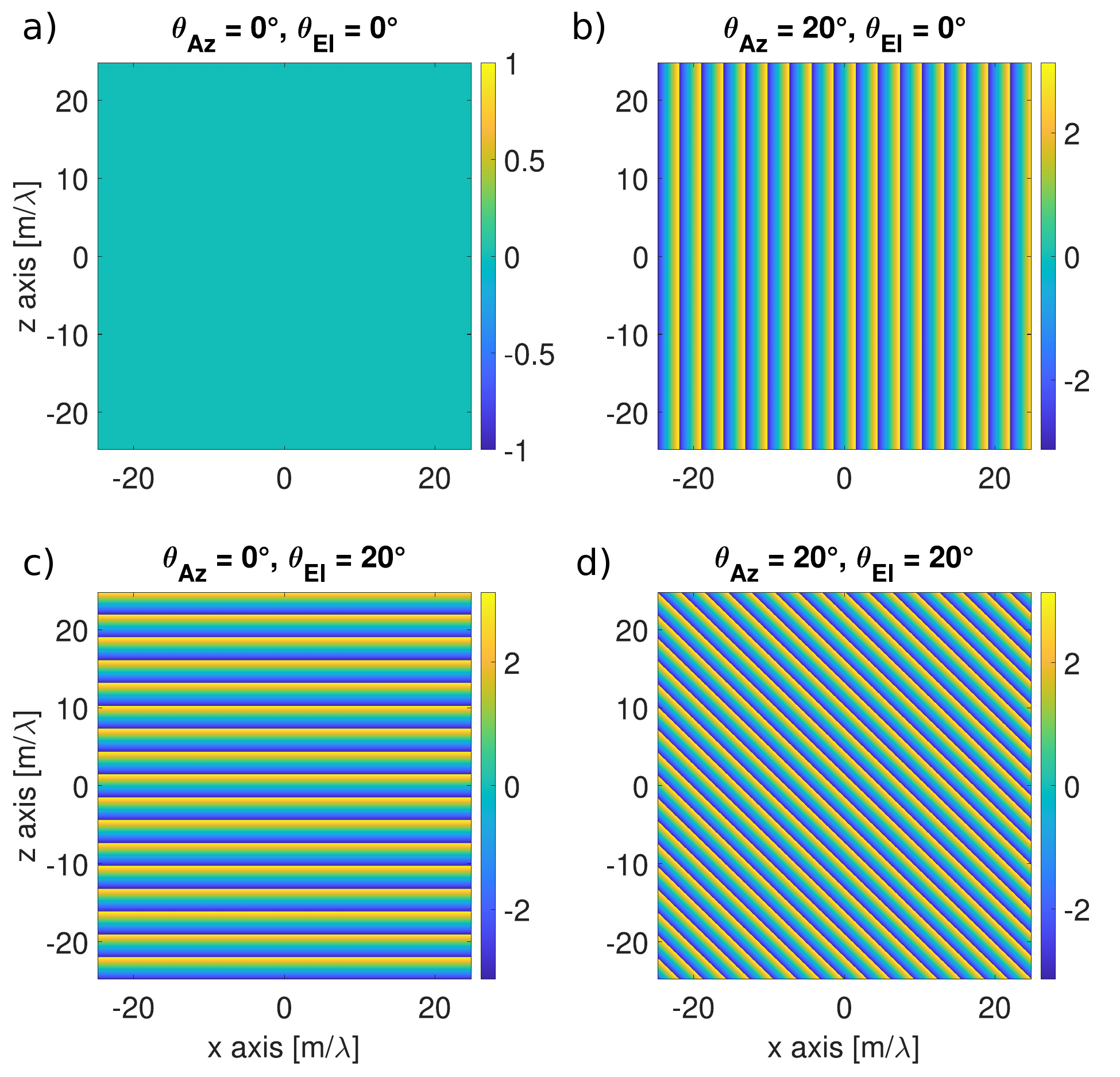}
    \caption{
    Phase distributions on antenna array surface for generation of Gaussian beams with varying $\theta_{Az}$ and $\theta_{El}$ angles.}
    \label{Fig:Gaussian_beam_phase_distribution}
\end{figure}

The phase wavefront of an unsteered Gaussian beam emitted from an antenna array on $xz$-plane can be described as a plane with equation $y = 0$~\cite{headland18}. The phase distribution on $xz$-plane generating the unsteered Gaussian beam is shown in Figure~\ref{Fig:Gaussian_beam_phase_distribution}(a). The steering of the Gaussian beam in an arbitrary direction is accomplished by describing the unsteered Gaussian beam function in the rotated coordinate system that corresponds to the steering direction. In the rotated coordinate system, the plane of the phase wavefront has the equation $y' = 0$. Reverting back to the original coordinate system using the transformation defined in (\ref{Eq:coordinates_transformation}), the equation of the tilted phase wavefront plane in the desired direction is given by:
\begin{equation}
y_{0} = x \, \tan{\theta_{Az}} + z \, \frac{\tan{\theta_{El}}}{\cos{\theta_{Az}}}. 
\end{equation} 

Solving the system of equation defined by (\ref{Eq:orthogonal line system of equations}) and utilizing the equation (\ref{Eq:d}), the minimal distance of the antenna element to the rotated phase wavefront is computed, along with the corresponding phase shifts (\ref{Eq:antennas_phase_shifts}). The phase shifts applied to each antenna element, as a function of the antenna element's coordinates in the $xz$-plane, are expressed as:
\begin{equation} \label{Eq:Gaussian beam phase distributaion equation}
\Delta \Phi (x_{a}, z_{a}) = k \, (x_{a} \cos{\theta_{El}} \, \sin{\theta_{Az}} + z_{a} \, \sin{\theta_{El}} ),
\end{equation}
which is a well-established equation for beam steering in the far field~\cite{balanis15b}, confirming the validity of the described steering approach.
Figures~\ref{Fig:Gaussian_beam_phase_distribution}(b), (c), and (d) illustrate the phase distributions derived using (\ref{Eq:Gaussian beam phase distributaion equation}) for various $\theta_{Az}$ and $\theta_{El}$ values.

\subsection{Bessel beam steering}

The unsteered Bessel beam's phase wavefront manifests as a cone in the positive domain of the $y$-axis~\cite{singh22c}, which is represented mathematically as follows:
\begin{equation}  \label{eq:cone}
    y =  \frac{h}{r} \, \sqrt{ (x^{2} + z^{2})},
\end{equation}
where $r$ is the radius of the cone at high $h$.
Steering the Bessel beam involves rotating the cone by $\theta_{Az}$ in azimuth and $\theta_{El}$ in elevation while maintaining the cone's geometry. In the equivalently rotated coordinate system ($x'$, $y'$, and $z'$), the equation of the cone is expressed as:
\begin{equation}  \label{Eq:cone}
     y' =  \frac{h}{r} \, \sqrt{ (x^{'2} + z^{'2})}.
\end{equation}
By utilizing the transformation defined by (\ref{Eq:coordinates_transformation}), we derive the equation of the rotated cone within the original coordinate system.

\begin{figure}[!htb]
    \centering
    \includegraphics[width=1\columnwidth]{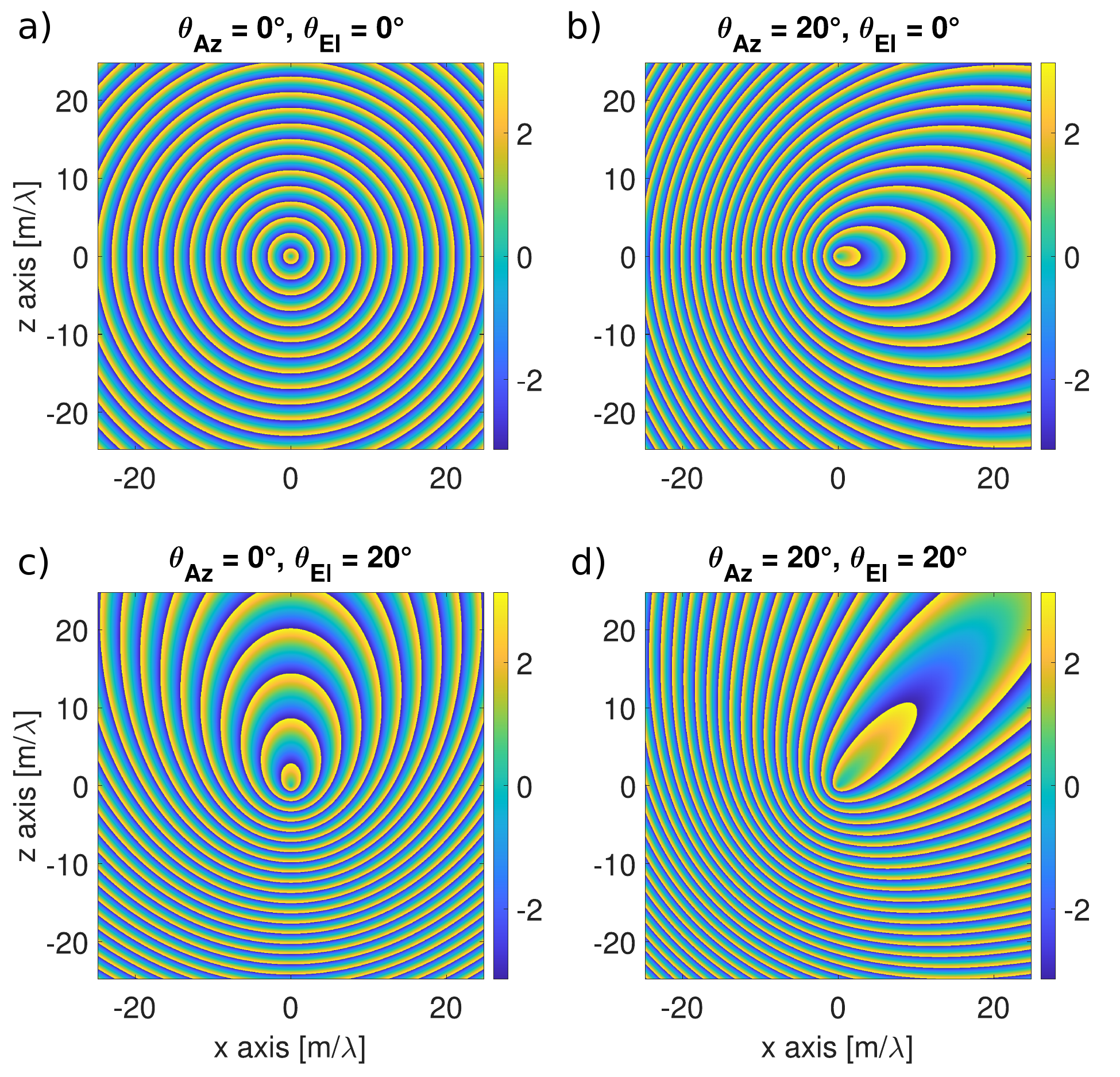}
    \caption{Phase distributions on antenna array surface for generation of Bessel beams with varying $\theta_{Az}$ and $\theta_{El}$ angles.}
    \label{Fig:Bessel beams phase distributions}
\end{figure}

Determining the minimum distance between the antenna element and the nearest point on the rotated phase wavefront involves solving the system of equations specified by (\ref{Eq:orthogonal line system of equations}) and utilizing equation (\ref{Eq:d}), as also depicted in Figure~\ref{Fig:Orthogonal lines: phase wavefronts - antennas}. The required phase shifts for antenna elements are calculated with (\ref{Eq:antennas_phase_shifts}). As the analytical solution to the system of equations~(\ref{Eq:orthogonal line system of equations}) was not found, the distances and, consequently, the phase progressions were calculated numerically in the Matlab software environment.
Figure~\ref{Fig:Bessel beams phase distributions} shows the acquired phase distributions on the $xz$-plane, which represent the phase shifts that must be applied to the antenna elements to produce the desired beam for various $\theta_{Az}$ and $\theta_{El}$ values.

\begin{figure}[!htb]
    \centering
    \includegraphics[width=1\columnwidth]{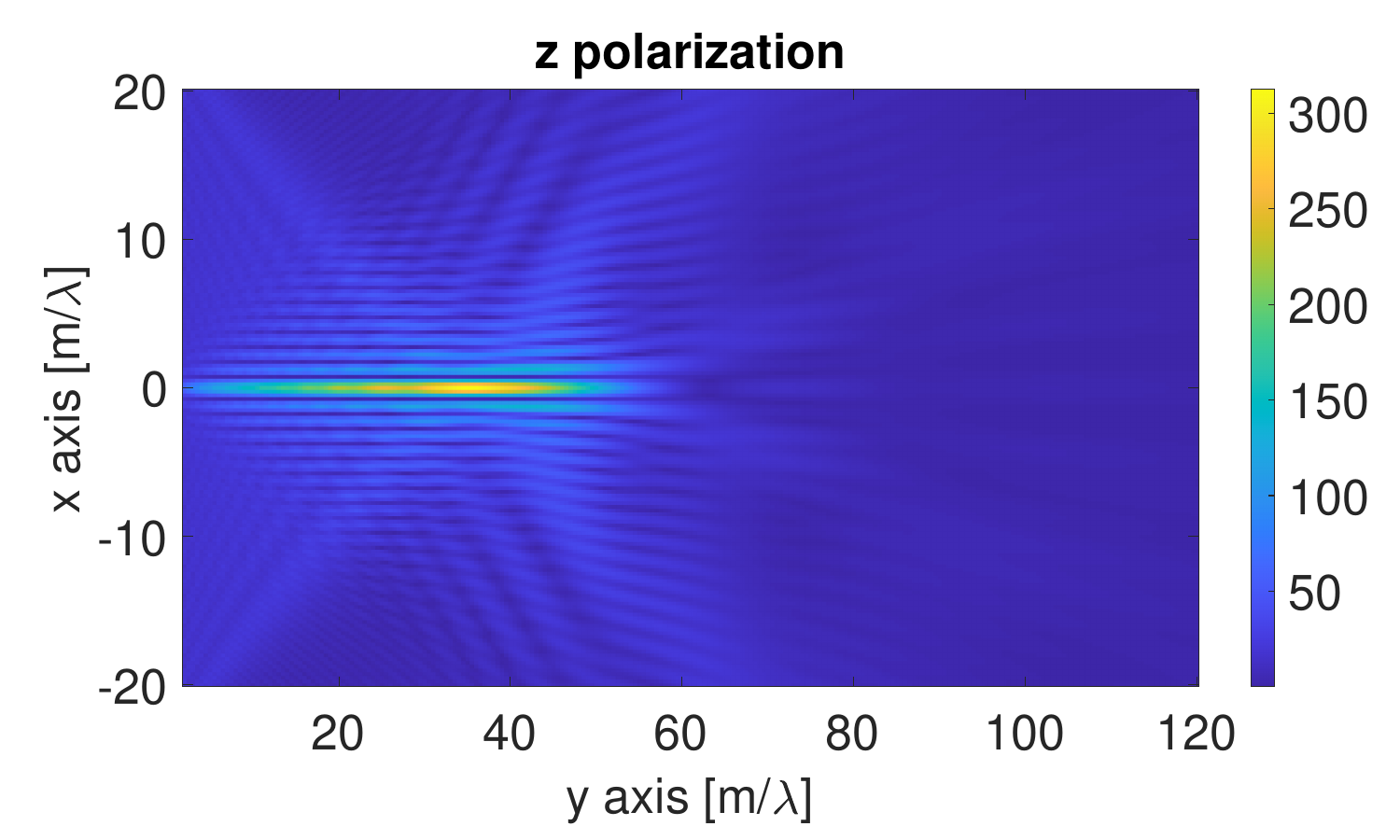}
    \caption{Bessel beam without steering.}
    \label{Fig:Unsteered Bessel beam}
\end{figure}

\begin{figure}[!htb]
    \centering
    \includegraphics[width=1\columnwidth]{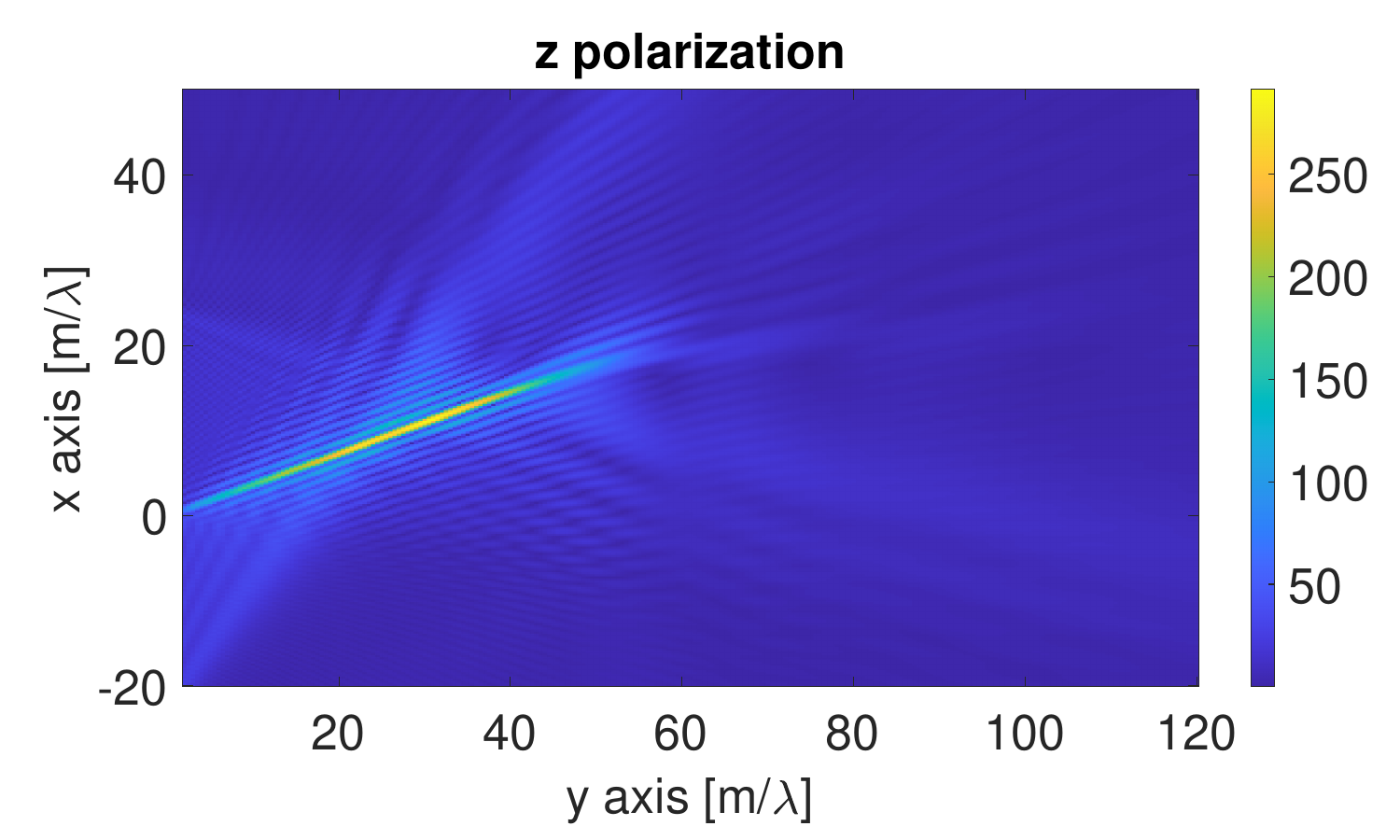}
    \caption{Bessel beam steered in azimuth, $\theta_{Az}$ = $20^{\circ}$.}
    \label{Fig:Bessel beam azimuth steering}
\end{figure}

\begin{figure}[!htb]
    \centering
    \includegraphics[width=1\columnwidth]{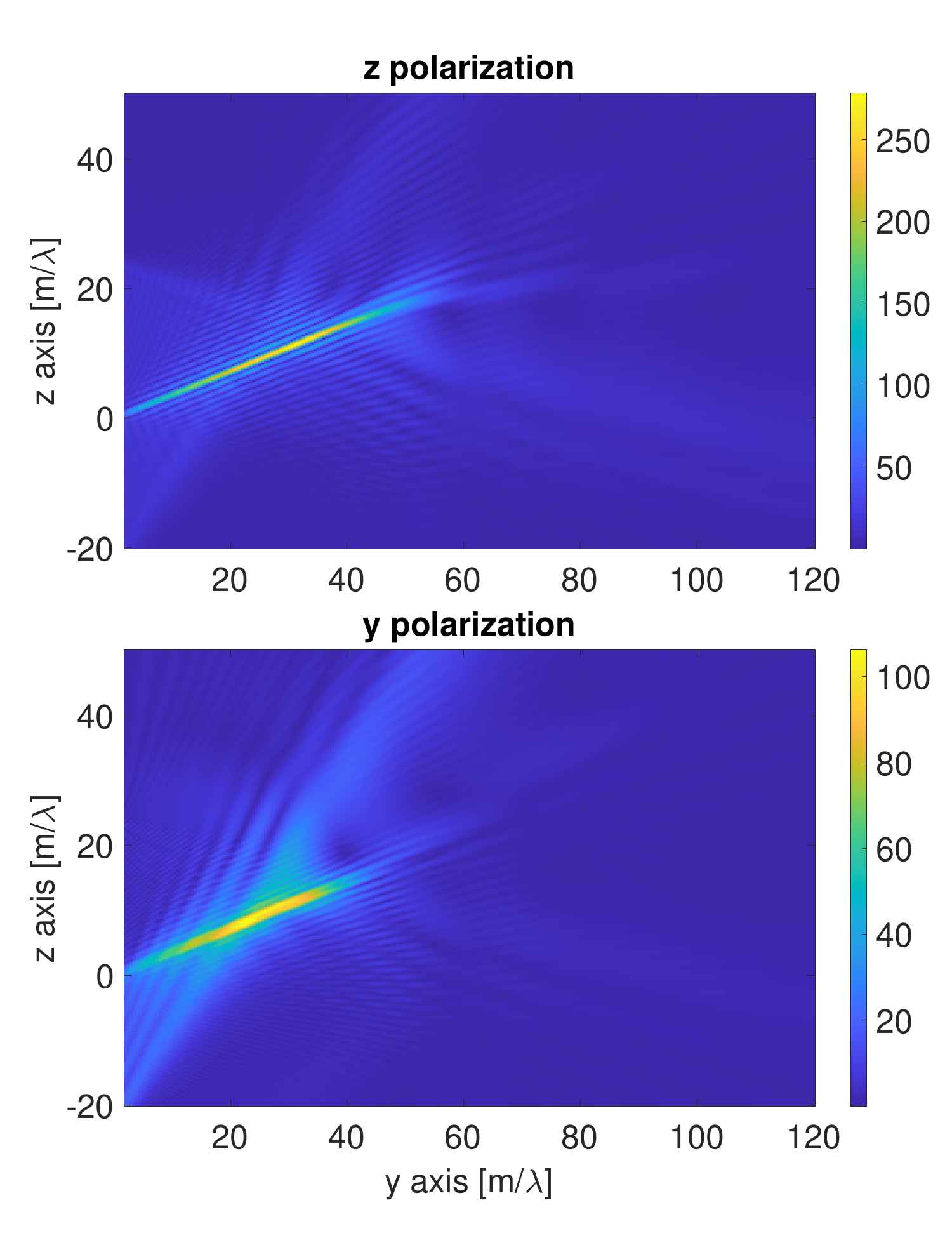}
    \caption{Bessel beam steered in elevation, $\theta_{El}$ = $20^{\circ}$.}
    \label{Fig:Bessel beam elevation steering}
\end{figure}

Figures~\ref{Fig:Unsteered Bessel beam},~\ref{Fig:Bessel beam azimuth steering}, and~\ref{Fig:Bessel beam elevation steering} show the Bessel beams generated by the phase distributions~\ref{Fig:Bessel beams phase distributions}(a), (b), and (c), respectively. The electric field of x-, y- and z-polarization was computed at each spatial point. We only plotted the electric field polarizations that do not have negligible electric field values. For the unsteered beam, only the z-polarization is shown, as the x- and y-polarizations have negligible values. When the beam is steered, the polarization mismatch is different when the beam is steered in azimuth and elevation. With azimuthal steering, only z-polarization is present, while steering in elevation results in significant y-polarization values. This presents a challenge since most receivers are designed to receive only one type of polarization, which reduces the power efficiency of the system.
The reason for this is that we have assumed that the polarization of the antenna elements is $\mathbf{u}_{\theta_{n}}$, a polarization associated with dipole antennas. The x-polarization contributions of the antenna elements to the electric field, calculated on the $xy$-plane or the $yz$-plane, cancel each other out. Similarly, the y-polarization contributions for points on the $xy$-plane cancel each other out. For this reason, only the z-polarization is significant if (1) the beam is not steered, since most of the beam is concentrated around the $y$-axis, or (2) the beam is steered only in azimuth since the beam is concentrated on the $xy$-plane. If the beam is steered in elevation, the contributions of the y-polarizations do not cancel each other out, so the y-polarization values are not negligible.
At points not on the $xy$- or $yz$-plane, all polarizations are present, with the influence of x- and/or y-polarization increasing with distance from the $y$-axis, further enhancing the effect of polarization mismatch. Consequently, if the beam is steered in both azimuth and elevation directions, all polarizations are expected, with the values of y-polarization increasing with increasing elevation angle and the values of x-polarization increasing with increasing azimuth angle.

The propagation range of Bessel beams is inherently limited by the finite transmit power and is a function of the antenna array's aperture size and the $h/r$ ratio~\cite{singh22c}.

\section{Conclusion}

This paper studies a method for steering beams with known phase wavefront functions towards arbitrary azimuth and elevation angles. The method involves rotation of the phase wavefront function by azimuth and elevation angle while maintaining the shape and estimation of the distance from the antenna elements to the nearest point on the rotated wavefront. The distances are proportional to the required phase shifts for the antenna element's excitation current to generate the steered beam.

Two examples of beam steering are presented. The first example utilized a Gaussian beam, while the second example focused on steering a Bessel beam. Phase distribution examples for various steering directions are considered. Additionally, for Bessel beam steering, the resulting beam shape is analysed, and its impact on polarization mismatch within the antenna array's near field is discussed. When the beam is steered only in azimuth there is no polarization mismatch assuming the receiver with z-polarization. However, steering the beam in elevation leads to non-negligible values of the cross-polarization, which has to be taken into account. When azimuth and elevation steering are used, all three polarizations exhibit non-negligible values, with varying magnitudes depending on the azimuth and elevation angles of the steering direction.

In future work, the proposed methodology can be extended to accommodate arbitrary radiation patterns, signals with higher bandwidth can be considered, and the potential signal distortions resulting from the steering process can be identified and studied.

\bibliographystyle{IEEEtran}
\bibliography{Beam_steering_used.bib}

\end{document}